\documentclass
[aps,amsmath,amssymb,longbibliography]{revtex4-2}%
\usepackage{amssymb}
\usepackage{graphicx}
\usepackage{amsmath}
\usepackage{amsfonts}%
\usepackage{siunitx}
\providecommand{\U}[1]{\protect\rule{.1in}{.1in}}
\usepackage{setspace}
\begin{document}

\title{Inclination-Induced Crossover in the Velocity Scaling of Lubrication-Mediated Droplet Motion}
\author{Haruka Hitomi and Ko Okumura\\Physics Department and Soft Matter Center, Ochanomizu University, 2-1-1
Ohtsuka, Bunkyo-ku, Tokyo 112-8610, Japan}
\date{\today}

\begin{abstract}

We experimentally investigate the creeping motion of water droplets sliding
along an inclined surface immersed in a viscous oil. The droplet velocity,
width, and height are measured over a broad range of droplet sizes and
inclination angles. The velocity increases with droplet radius according to a
power law, but the corresponding exponent depends strongly on the inclination
angle. At small inclinations, the exponent is close to 3/2, whereas at larger
inclinations it progressively approaches 9/4. The results therefore reveal a
continuous inclination-induced crossover in the velocity-size scaling of
lubrication-mediated droplet motion. A scaling analysis based on
Landau-Levich-Derjaguin film formation and viscous dissipation in the dynamic
meniscus reproduces these limiting behaviors, which correspond to the
quasi-spherical and pancake scaling regimes, respectively.

Remarkably, the evolution of the velocity scaling is substantially more
pronounced than the corresponding evolution of the macroscopic droplet shape.
While the velocity data progressively approach the pancake scaling law at
large inclinations, the global droplet dimensions evolve in a more complex
manner than expected from a simple pancake-shape picture. In particular, the
approach to pancake-like velocity scaling occurs even when the measured
droplet dimensions remain far from the corresponding pancake-limit geometry.

These observations suggest a crossover in the dominant lubrication-dissipation
mechanism beneath the droplet that is not directly reflected in the global
droplet morphology. The results identify inclination angle as a key control
parameter governing lubrication-mediated droplet motion and highlight the
distinction between global shape evolution and local dissipation dynamics in
liquid-liquid systems.

\end{abstract}
\maketitle


\section{Introduction}

The motion of a liquid drop creeping on an inclined plate within an immiscible
viscous medium is quite complex. This complexity arises from the presence of a
thin lubricating film formed between the drop and the slope, the latter of
which is fully wetted by the surrounding medium. To understand the dynamics,
we must simultaneously account for two aspects: the viscous dissipation within
the thin film and the deformation of the drop shape by gravity.

The physics of thin-film formation has long been a subject of active
investigation. Fundamental theories for film thickness were first established
for flat plates by Landau, Levich, and Derjaguin \cite{levich1942,
derjaguin1943, derjaguin1993}, and then for cylindrical tubes by Bretherton
\cite{Bretherton}. While these theories provide a robust foundation for
predicting film thickness, the terminal velocity of a creeping drop is further
governed by its overall shape.

Unlike a rigid solid sphere rolling down a dry slope, a deformable drop
supported by a thin lubricating film exhibits fundamentally different rolling
and sliding dynamics. For instance, Bico et al.~demonstrated that a rigid
sphere migrating on a liquid-coated incline exhibits non-trivial scaling
exponents due to the capillary-viscous coupling within the wetting meniscus
\cite{bico2009}. For a deformable fluid drop, the physical picture becomes
even more intricate due to the interplay between gravity-induced shape
deformation and film-mediated lubrication. Mahadevan and Pomeau theoretically
demonstrated that, for a small quasi-spherical drop with a flattened bottom
whose radius is smaller than the capillary length, the velocity varies
inversely with the drop radius \cite{mahadevan1999rolling}. This prediction is
experimentally confirmed \cite{richard1999viscous}. In contrast, Aussillous
and Qu{\'{e}}r{\'{e}} experimentally demonstrated that for a bubble creeping
beneath an inclined slope within a viscous oil, the velocity increases with
the bubble radius \cite{aussillous2002bubbles}. This behavior arises because
the motion is dominated by viscous dissipation within the dynamic meniscus of
the thin lubricating film. Following this seminal work, research on sliding
bubbles has expanded to cover various physical conditions, including
moderate-to-high Reynolds numbers \cite{dubois2016, barbosa2019}, transient
bouncing dynamics \cite{esmaili2019}, and interfacial electrolyte effects
\cite{del2011inhibition}.

Theoretically, Hodges et al. first showed that the drop size and viscosity
ratio govern the creeping motion of a drop on a gently inclined plane
\cite{hodges2004motion}. Subsequent numerical studies explored complementary
aspects of this problem. Lattice Boltzmann simulations were employed to
quantify rolling-to-sliding transitions in moving droplets \cite{thampi2015}.
In parallel, boundary-integral simulations revealed how droplet deformation,
lubrication-film formation, and hydrodynamic wall interactions jointly
determine droplet velocity and droplet-wall separation under
low-Reynolds-number conditions \cite{Griggs2007}. More recently,
high-resolution three-dimensional boundary-integral simulations have provided
detailed predictions of thin-film structure and droplet dynamics over a broad
range of Bond numbers and viscosity ratios \cite{zinchenko2024}. Recently,
this physical picture has expanded to encompass engineered lubricating
interfaces \cite{daniel2017oleoplaning, keiser2017}, viscosity-dependent
rolling regimes of droplets in viscous media \cite{PhysRevFluids.3.023601},
sliding-to-rolling transitions of yield-stress fluids \cite{carneri2023}, and
anomalous viscosity-enhanced transport in confined capillaries
\cite{vuckovac2020viscosity}.

Despite these advances, it remains unclear how the scaling law governing the
motion of deformable liquid drops evolves with inclination angle in
lubrication-mediated liquid-liquid systems. To address this question, we
investigate the motion of water drops sliding along an inclined plane
submerged in an immiscible viscous oil. Our primary focus is on how the
scaling behavior changes with inclination angle. Drops with a radius larger
than the capillary length are often referred to as `pancakes' due to their
flattened shape \cite{mahadevan1999rolling, aussillous2002bubbles,
hodges2004motion}. Although the distinction between quasi-spherical and
pancake droplets is conventionally discussed in terms of droplet size, the
corresponding consequences for lubrication-mediated motion remain less clear.
Our objective is not to identify a direct geometric transition of droplet
shape, but rather to determine how the scaling law governing
lubrication-mediated motion evolves with inclination angle.

\section{Experimental}

\begin{figure}[tbh]
\centering\includegraphics[width=\textwidth]{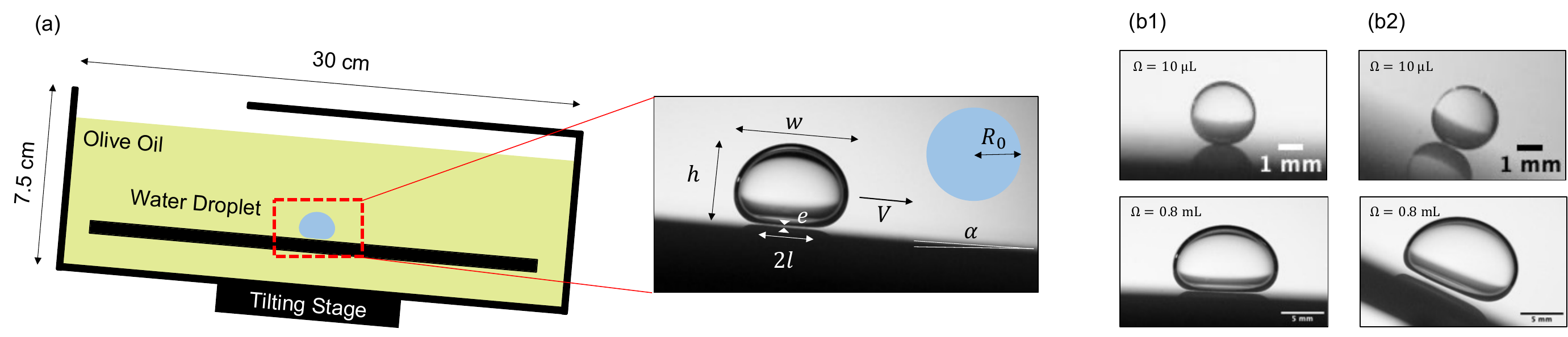}\caption{(a) Left:
Schematic of the experimental setup with a slope $\alpha$. Right: Photograph
of a water drop (volume $\Omega$ = 400 $\mu$L) creeping down at velocity $V$
on a slope with $\alpha$ = 5$^{\circ}$. (b1) Small and large drops on a slope
of $\alpha$ = 2$^{\circ}$. (b2) The same droplets on a steeper slope. The
increase in inclination leads to a noticeable increase in both width and
height, accompanied by enhanced front--rear asymmetry, particularly for the
larger drop. The images in (b1) and (b2) are shown at the same scale.}%
\label{Fig:setup}%
\end{figure}As illustrated in Fig.~\ref{Fig:setup} (a), an acrylic container
(30 cm in length, 7.5 cm\ in height and 5 cm in width) equipped with a
27-cm-long slope was filled with olive oil (FUJIFILM Wako Pure Chemical
Corporation, $\rho_{o}$ =910 kg/m$^{3}$). Since the oil viscosity $\eta$
depends on temperature, we monitored the temperature during each trial and
calibrated the viscosity using a rheometer (MCR 302, Anton Paar). The
resulting values ranged between 60 and 70 mPa$\cdot$s, which was reflected in
the ensuing analysis. The water-oil interfacial tension was measured by
pendant drop method ($\gamma=27.0\pm1.6$\% mN/m). The density difference
$\Delta\rho$ between the olive oil and water was 90 kg/m$^{3}$, which yielded
a capillary length, defined by $\kappa^{-1}=\sqrt{\gamma/(\Delta\rho g)}$, of
$5.5\pm1.8$\% mm, where $g$ is the gravitational acceleration. The entire
container was inclined to set the slope angle $\alpha$ from 0.4$^{\circ}$ to
25$^{\circ}$ with an accuracy of $\pm$ 0.3$^{\circ}$. Water drops of
controlled volume ($\Omega=10,20,40,100,200,400,800,1400,\text{ and
}3000\,\mathrm{\mu L}$) were deposited onto the slope using micropipettes
(Pipette-Guy for 10 to 800 $\mu$L with a precision $\pm1\%$; Eppendorf for
1400 and 3000 $\mu$L with a precision $\pm0.6\%$). The equivalent radius of
the undistorted droplet, $R_{0}$, calculated from $\Omega=\frac{4}{3}\pi
R_{0}^{3}$, ranged from $1.36$ to $8.92\,$\ mm. The uncertainty in $R_{0}$
resulting from volume measurement was below 0.3$\%$ throughout the
investigated range. The movement of the drop was recorded with a digital
camera (Nikon D800E) with a lens (AF-S Micro NIKKOR 60mm 1:2.8G ED). The
extracted images were analyzed using ImageJ. We measured the maximum drop
width $w$ (streamwise length along the slope), the maximum height $h$
(measured perpendicular to the slope), and the creeping velocity $V$ by
tracking the leading edge of the drop. Beneath the moving drop, a thin film of
olive oil (thickness $e$, contact radius $l$) was formed, which implies that
olive oil completely wetted the slope and the drop did not stick to the surface.

\section{Experimental Results\label{Experimental Results}}

\begin{figure}[tbh]
\centering
\includegraphics[width=0.9 \textwidth]{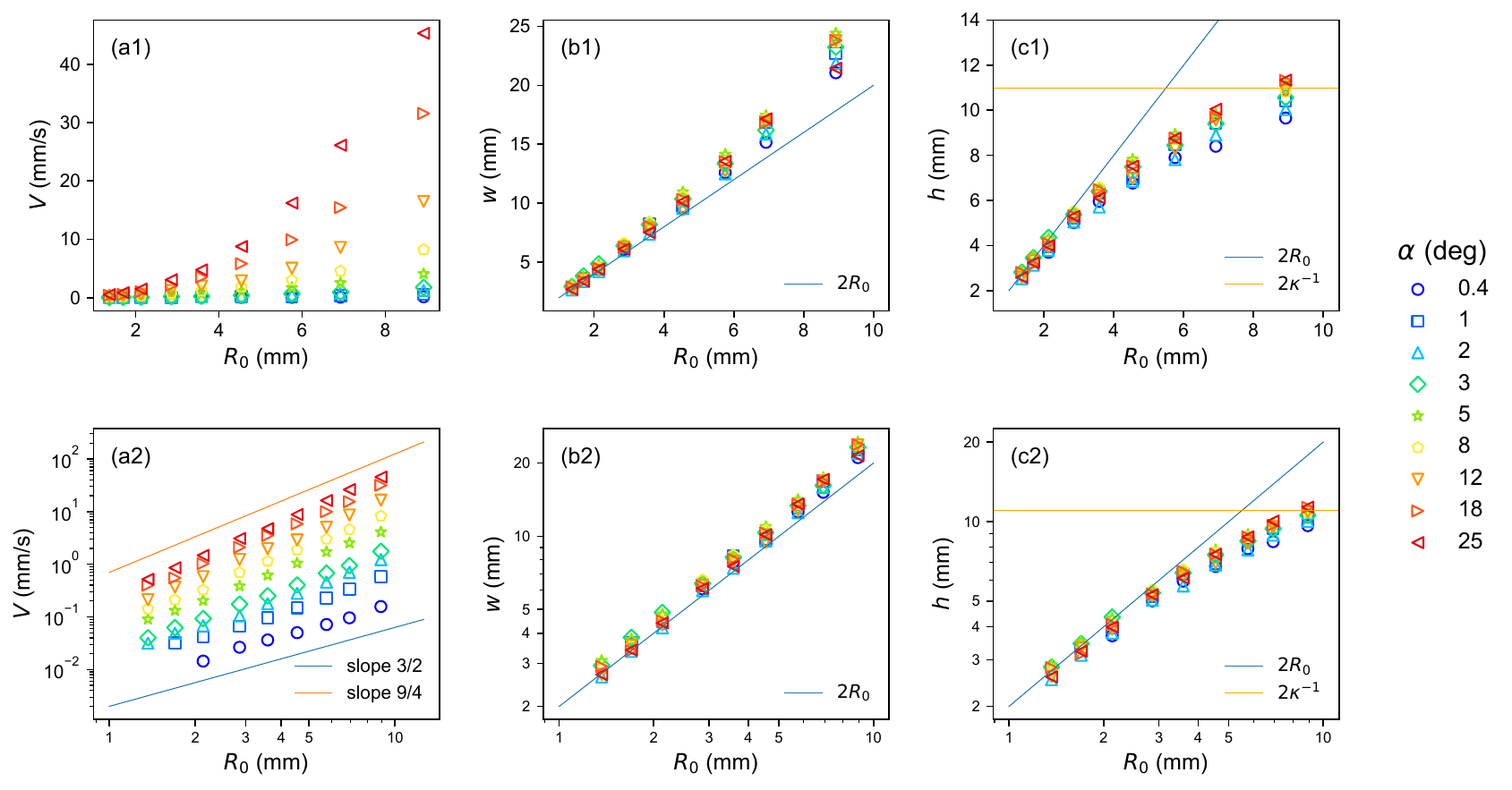} \caption{ Droplet velocity
$V$ (a1, a2), maximum width $w$ (b1, b2), and maximum height $h$ (c1, c2) as a
function of the equivalent drop radius $R_{0}$ for various inclination angles
$\alpha$. The top row (a1--c1) represents linear scales, and the bottom row
(a2--c2) represents log-log scales. In (a2), guidelines representing power-law
slopes of 3/2 and 9/4 are plotted. In (b1, b2) and (c1, c2), the solid
guidelines represent the diameter of an undistorted spherical drop ($2R_{0}$)
and the theoretical pancake limit ($2\kappa^{-1}$), respectively. }%
\label{Fig:results}%
\end{figure}

In Fig.~\ref{Fig:results}, we show the velocity $V$, drop width $w$ and height
$h$ as a function of the equivalent drop radius $R_{0}$ for various angles
$\alpha$, on both linear (top: a1 to c1) and log-log (bottom: a2 to c2) scales.

As for the velocity, in the Fig.~\ref{Fig:results} (a), we find that the
velocity $V$ increases with $R_{0}$ at each fixed $\alpha$, exhibiting
power-law behavior. The exponent (slope) increases from approximately 3/2 to
9/4 as the angle $\alpha$ increases.

Regarding the drop shape at a fixed $\alpha$, Fig.~\ref{Fig:results} (b) and
(c) show the flattening of the drop with increasing size. Both $w$ and $h$ are
close to $2R_{0}$ (the value expected for a spherical drop) at small $R_{0}$,
but they respectively increase and decrease from the spherical value as
$R_{0}$ increases, reflecting the flattening. In addition, $h$ tends to
saturate to 2$\kappa^{-1}$, the theoretically expected static puddle limit for
sufficiently large droplets \cite{de2013capillarity}$.$

Importantly, for a fixed $R_{0}$, the inclination angle affects the shape in a
non-trivial manner. Because the velocity $V$ increases with $\alpha$ for a
fixed $R_{0}$, this complex behavior can be interpreted as a dynamic effect.
While $w$ increases with $\alpha$---suggesting that the drop flattens along
the slope---the maximum height $h$ also exhibits a systematic increase. This
tendency is already visible qualitatively in Fig.~\ref{Fig:setup}(b1, b2),
particularly for the larger droplet. Since the total droplet volume $\Omega$
is conserved ($\Omega\sim ww^{\prime}h$), this concurrent increase in both $w$
and $h$ mathematically implies that the transverse width $w^{\prime}%
$---measured in the direction perpendicular to both the rolling motion and the
slope normal---must decrease with $\alpha$. While $w^{\prime}$ was not
directly measured, visual inspection of the droplets was consistent with such
a decrease.

To further clarify the droplet deformation, we examine the saturation of the
maximum height $h$ with increasing droplet size at fixed inclination angle.
While the saturation of $h$ toward the static puddle limit of $2\kappa^{-1}$
for sufficiently large drops is physically expected from a hydrostatic
perspective \cite{de2013capillarity}, the observation that this saturation is
systematically delayed to larger drop sizes at steeper inclinations is highly
non-trivial. Moreover, at a fixed equivalent radius $R_{0}$, the maximum
height $h$ remarkably increases with $\alpha$. This behavior may be
interpreted in terms of two contributing mechanisms. First, as the inclination
angle $\alpha$ increases, the effective gravitational component compressing
the drop normal to the slope, $g\cos\alpha$, decreases. This reduction in the
normal compressive force allows the interfacial tension to counteract the
gravitational flattening more effectively, prompting a "surface-tension
recoil" toward its quasi-spherical state and thereby delaying the onset of
flattening. Second, the misalignment between the vertical gravity vector and
the normal force from the substrate, combined with the viscous drag within the
lubricating oil film, induces a pronounced fore-aft asymmetry of the droplet
profile, as qualitatively illustrated in Fig.~\ref{Fig:setup}(b2). Because our
measured $h$ is defined as the maximum height normal to the slope, it is
sensitive to this asymmetry, which contributes to the observed increase in $h$
and shifts the apparent saturation point toward larger $R_{0}$ at steeper slopes.

Finally, the experimentally accessible range of droplet sizes is limited by
the onset of motion. At sufficiently small inclination angles and droplet
volumes, the gravitational driving force is unable to overcome the sticking
threshold of the system. Consequently, the 10 and 20 $\mathrm{\mu L}$ droplets
at = 0.4$^{\circ}$ and the 10 $\mathrm{\mu L}$ droplet at = 1$^{\circ}$
remained stationary, and no velocity measurements could be obtained for these conditions.

\section{Theory}

\subsection{Static shape and characteristic length scales}

\begin{figure}[tbh]
\centering\includegraphics[width=0.5\textwidth]{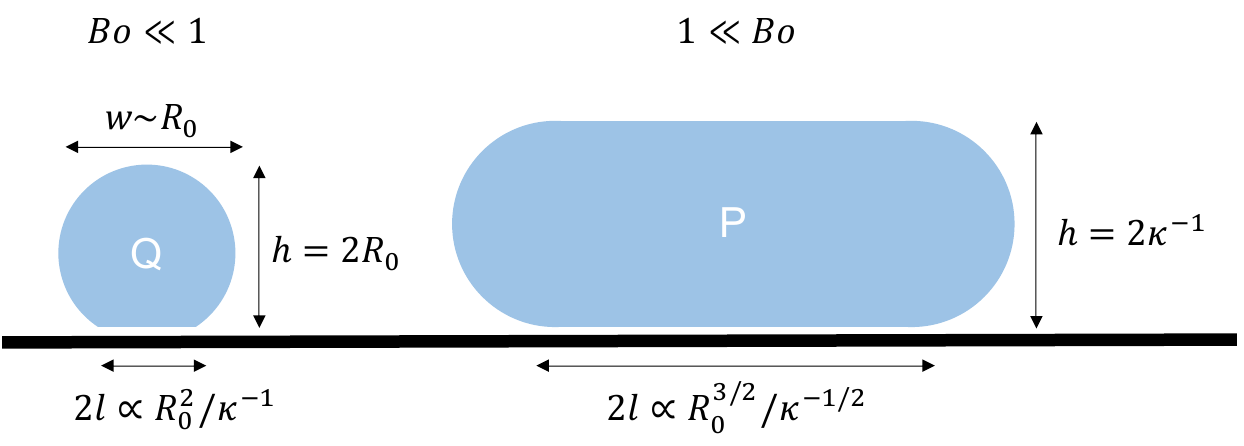} \caption{
Schematic illustration of the droplet shapes and their characteristic scaling
parameters in the two regimes: the quasi-spherical drop (Q) in the small Bond
number limit ($Bo\ll1$) and the pancake drop (P) in the large Bond number
limit ($1\ll Bo$). The parameters $h$, $w$, and $2l$ denote the maximum
height, maximum width, and contact disk diameter, respectively, with their
scaling relations in terms of the equivalent radius $R_{0}$ and the capillary
length $\kappa^{-1}$. }%
\label{Fig:shape}%
\end{figure}

By considering the small $\alpha$ limit, in which the velocity $V$ is zero,
and thus the shape of a drop on a surface is simply determined by the balance
between gravity and capillary forces, drop shapes are classified into two
regimes (Fig.~\ref{Fig:shape}): Pancake (a flat drop with saturated height, $1
\ll Bo$) and Quasi-sphere (a drop with a distinct flat spot, $Bo \ll1$). They
are characterized by the capillary length $\kappa^{-1} = \sqrt{\gamma/
(\Delta\rho g)}$ and the Bond number $Bo = \Delta\rho g R_{0}^{2} / \gamma=
(R_{0} / \kappa^{-1})^{2}$.

The pancake drop is significantly flattened by gravity. Given the complete
wetting condition, the height is fixed by gravity and given as $h=2\kappa
^{-1}$ \cite{de2013capillarity, okumura2003waterspring}. Volume conservation
($4\pi R_{0}^{3}/3\sim2\pi l^{2}\kappa^{-1}$, which becomes exact in the limit
$\kappa^{-1}\ll l$) yields an estimate for the contact size $l$, given by
$l=\sqrt{2/3}R_{0}(R_{0}/\kappa^{-1})^{\beta}$ with $\beta=1/2$.

For a quasi-spherical drop, while capillary forces still dominate, gravity
causes a slight deformation and creates a contact disk of radius $l$. This
flattening causes a change in the surface area, involving both a decrease and
an increase. Firstly, there is a reduction in the area of the spherical cap
scaled as $l^{4} / R_{0}^{2}$. Secondly, there is an increase in the surface
area of the bulk of the drop to conserve volume. Noting that the center of
mass is lowered by $\delta$ and using the geometric relation $l^{2} \sim R_{0}
\delta$, the volume of the deformed cap scales as $\sim l^{2} \delta\sim l^{4}
/ R_{0}$. When the change in the drop radius is $\Delta R_{0}$, the
corresponding volume change is $R_{0}^{2} \Delta R_{0} \sim l^{4} / R_{0}$,
which yields an area increase of $R_{0} \Delta R_{0} \sim l^{4} / R_{0}^{2}$.
By balancing these two surface energy contributions against the gravitational
potential energy, we find that the radius of the flat contact disk scales as
$l \sim R_{0} (R_{0} / \kappa^{-1})^{\beta}$ with $\beta= 1$
\cite{mahadevan1999rolling, hodges2004sliding}. In summary, the scale $l$ is
given by:%

\begin{equation}
l\sim R_{0}\left(  \frac{R_{0}}{\kappa^{-1}}\right)  ^{\beta}\quad
\text{with}\quad\beta=%
\begin{cases}
1/2 & \text{for a pancake drop}\\
1 & \text{for a quasi-spherical drop}%
\end{cases}
\label{eq:l}%
\end{equation}

\subsection{Scaling Laws}

To describe the creeping dynamics of pancake and quasi-spherical drops, we
note the existence of a thin film of olive oil formed between the drop and the
slope surface. We assume that the film thickness $e$ is governed by the
Landau-Levich-Derjaguin (LLD) theory, because we focus on the region
$Ca^{1/3}\ll1$, in which this law is valid \cite{levich1942}. Within this
film, a velocity gradient $\sim V/e$ develops specifically in the dynamic
meniscus region, characterized by the length scale $\lambda$. The LLD theory
is based on the Stokes equation and matching condition:%

\begin{equation}
\eta\frac{V}{e^{2}} \sim\frac{1}{\lambda}\left(  \frac{\gamma}{h}\right)  ,
\quad\frac{1}{h} \sim\frac{e}{\lambda^{2}} \label{eq:lld}%
\end{equation}

In the first relation of Eq.~\eqref{eq:lld}, the viscous stress balances with
the pressure gradient originating from Laplace's pressure jump of the order of
$\gamma/h$. This is because the height $h$, rather than the radius $R_{0}$, is
the characteristic length scale of the curvature for quasi-spherical drops, as
well as for pancake drops. For the former, $h$ simply scales as $R_{0}$,
while, for the latter, the first and second radii of curvature scale as $h$
and $R_{0}$, and thus the curvature scales as $1/h$ because $h\ll R_{0}$. The
second relation in Eq.~\eqref{eq:lld} is the matching condition, where the
curvature $e/\lambda^{2}$ defined by the dynamic meniscus length $\lambda$
matches the static curvature $1/h$. Solving Eq.~\eqref{eq:lld} for $\lambda$
and $e$ yields the following scaling laws:%

\begin{equation}
\lambda\sim h\left(  \frac{\eta V}{\gamma}\right)  ^{1/3},\quad e\sim h\left(
\frac{\eta V}{\gamma}\right)  ^{2/3} \label{eq:lld2}%
\end{equation}

Assuming that viscous dissipation is dominated by that in the dynamic meniscus
\cite{aussillous2002bubbles}, the viscous force scales as $F_{v}\sim
\eta(V/e)\lambda l\sim\gamma lCa^{2/3}$, where $Ca=\eta V/\gamma$ is the
capillary number. Balancing $F_{v}$ with the gravitational driving force $\rho
gR_{0}^{3}\sin\alpha$ and substituting $l$ from Eq.~\eqref{eq:l}, we obtain
the scaling law for the normalized velocity with $\tilde{\beta}=\frac{3}%
{2}(2-\beta)$:
\begin{equation}
Ca\sim\left(  \frac{R_{0}}{\kappa^{-1}}\right)  ^{\tilde{\beta}}\sin^{\frac
{3}{2}}\alpha\quad\text{with}\quad\tilde{\beta}=%
\begin{cases}
9/4 & \text{for a pancake drop}\\
3/2 & \text{for a quasi-spherical drop }%
\end{cases}
\label{eq:ca}%
\end{equation}

\section{Experiment and Theory}

\begin{figure}[tbh]
\centering
\includegraphics[width=0.9 \textwidth]{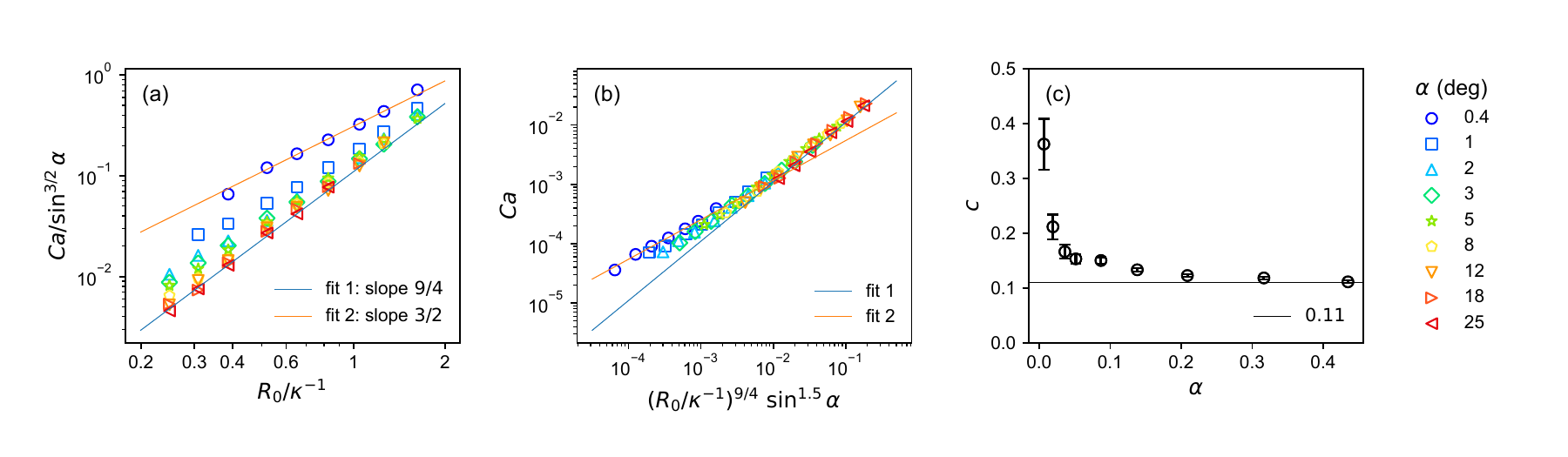} \caption{ (a) $Ca/\sin
^{3/2}\alpha$ versus $R_{0}/\kappa^{-1}$ on log-log scales. (b) $Ca$ versus
$(R_{0}/\kappa^{-1})^{9/4}\sin^{3/2}\alpha$ on log-log scales. (c) Numerical
pre-factor $c$ versus $\alpha$ on linear scales. }%
\label{Fig:slope}%
\end{figure}

Fig.~\ref{Fig:slope}(a) displays $Ca/\sin^{3/2}\alpha$ as a function of the
normalized drop radius $R_{0}/\kappa^{-1}$, where all the data satisfy the
condition $Ca^{1/3}<0.3$. This condition confirms that the LLD theory is
indeed valid in this regime. Note that, based on Eq.~\eqref{eq:ca}, the
scaling exponent $\tilde{\beta}$ of $R_{0}/\kappa^{-1}$ is predicted to be
$3/2$ for the quasi-spherical regime ($\beta=1$) and $9/4$ for the pancake
regime ($\beta=1/2$).

As seen in Fig.~\ref{Fig:slope}(a), at the smallest angle $\alpha=0.4^{\circ}%
$, the exponent is close to the quasi-spherical value of $3/2$. However, the
observed exponent for each set of parameters systematically increases with the
inclination angle $\alpha$, deviating from the quasi-spherical value of $3/2$
and saturating toward the pancake value of $9/4$. To confirm this dynamic
transition, $Ca$ is plotted against $(R_{0}/\kappa^{-1})^{9/4}\sin^{3/2}%
\alpha$ in Fig.~\ref{Fig:slope}(b). For larger $\alpha$ (especially $8^{\circ
}$ to $25^{\circ}$), the data are reasonably described by a single master
curve over more than one decade, which suggests that the droplets on steeper
slopes are governed by pancake-like scaling. The evolution with inclination
angle is further quantified in Fig.~\ref{Fig:slope}(c). By fitting the data to
the scaling law $Ca=c(R_{0}/\kappa^{-1})^{9/4}\sin^{3/2}\alpha$, the
pre-factor $c$ is found to saturate at 0.11. This behavior is fully consistent
with Fig. \ref{Fig:slope}(a), suggesting a progressive approach toward
pancake-like lubrication scaling at higher inclination angles, even while the
overall macroscopic droplet profile remains transitional.

\section{Discussion}

A central observation of the present study is that the scaling behavior
governing droplet motion evolves systematically with increasing inclination
angle. As shown in Fig. 4, the exponent relating the capillary number to
droplet size increases from a value close to the quasi-spherical prediction of
3/2 toward the pancake prediction of 9/4. The collapse of the high-inclination
data using the pancake scaling law further suggests that the lubrication
dynamics progressively approach the pancake regime as the inclination angle increases.

By contrast, the macroscopic droplet dimensions shown in
Fig.~\ref{Fig:results}(b) and (c) evolve in a considerably more complex
manner. Most of the data that are reasonably described by the pancake scaling
law of Eq.~\eqref{eq:ca} still remain far from the pancake-limit height
$h=2\kappa^{-1}$. Thus, although the velocity scaling approaches the pancake
prediction at large inclination angles, the corresponding macroscopic shape
does not exhibit an equally clear approach to the pancake limit.

This discrepancy is particularly evident in the behavior of the droplet width.
A naive interpretation might suggest that, for droplets approaching the
pancake limit, the measured width $w$ would reflect the scaling of the
characteristic contact size $l$. Yet even when the velocity scaling has
already approached the pancake prediction, the observed dependence of $w$ on
$R_{0}$ remains substantially different from the pancake scaling predicted by
Eq.~\eqref{eq:l}. At the same time, volume conservation together with the
observed increases in both $w$ and $h$ implies a reduction of the transverse
width $w^{\prime}$ as the inclination angle increases. These observations
indicate that the global droplet dimensions respond to inclination in a more
complex manner than expected from a simple pancake-shape picture.

The results therefore suggest that the evolution of the dissipation mechanism
and that of the global droplet morphology are not synchronized. While the
velocity progressively approaches the pancake scaling regime with increasing
inclination angle, the macroscopic droplet shape remains transitional and
exhibits more complicated behavior. Since viscous dissipation is concentrated
within the dynamic meniscus, the terminal velocity can reflect changes in the
local lubrication geometry even when the overall droplet shape remains
transitional. The present observations therefore suggest that a crossover in
lubrication dynamics may occur before a comparable crossover becomes apparent
in the global droplet morphology.

Such a separation between global morphology and dynamics is consistent with
previous boundary-integral studies demonstrating the strong influence of
lubrication films on droplet motion and hydrodynamic wall interactions
\cite{Griggs2007, zinchenko2024}. Accordingly, the local lubrication region
responsible for viscous dissipation appears to undergo a dynamical crossover
that is not directly reflected in the overall droplet geometry. The emergence
of such a simple scaling law despite the complexity of the macroscopic
geometry is reminiscent of situations in which robust scaling behavior emerges
from geometrically complex interfacial confinement \cite{hitomi2025}.

The present experiments do not directly resolve the contact radius $l$, which
is challenging because the lubrication-film thickness varies continuously
across the contact region, making the effective contact radius difficult to
define unambiguously. Consequently, we infer a crossover of the effective
contact-length scaling from the velocity data rather than directly observing
it. The proposed evolution of the lubrication-region scaling should therefore
be regarded as an interpretation supported by the measured dynamics rather
than a direct measurement of the contact region itself.

A further feature of the data is the saturation of the numerical prefactor $c$
at approximately 0.11 for sufficiently large inclination angles, as shown in
Fig. 4(c). The emergence of a common prefactor accompanies the collapse of the
high-inclination data onto the pancake scaling law, suggesting that the
dominant dissipation mechanism becomes relatively insensitive to further
increases in inclination once this regime is reached. However, the magnitude
of the prefactor remains significantly smaller than previously reported
values. Specifically, the saturated value $c\simeq0.11$ is approximately three
times smaller than the coefficient $c\simeq0.34$ inferred from the bubble
experiments \cite{aussillous2002bubbles} and more than four times smaller than
the theoretical prediction $c\simeq0.474$ derived for creeping droplets on
gently inclined surfaces \cite{hodges2004sliding}.

The origin of this discrepancy remains unclear. One possibility is that the
detailed structure of the lubrication film differs from that assumed in
existing models. In addition, interfacial mobility, finite deformation of the
moving droplet, or other sources of viscous dissipation may alter the
numerical prefactor without changing the observed scaling exponents. Direct
measurements of the lubrication film thickness and contact geometry would
therefore provide a challenging but important test of the present interpretation.

Alternative explanations appear less consistent with the experimental
observations. Although the largest droplets occupied a substantial fraction of
the container width, no systematic reduction of velocity associated with
geometric confinement was observed. Likewise, balancing gravity with
conventional Stokes drag $F_{S}\sim\eta VR_{0}$ would predict $Ca\sim\left(
\frac{R_{0}}{\kappa^{-1}}\right)  ^{2}\sin\alpha$, which differs from the
measured dependence on $\sin^{3/2}\alpha$. These observations support the
interpretation that lubrication-induced dissipation dominates the droplet
dynamics in the present system. Finally, the applicability of the
Landau-Levich-Derjaguin framework appears reasonable throughout most of the
investigated parameter range. The experiments were conducted predominantly
within the regime $Ca^{1/3}<1$, where the classical lubrication scaling is
expected to provide the leading-order description of film formation.

Taken together, the results suggest that inclination angle acts as more than a
simple driving parameter. In addition to determining the gravitational force
along the slope, it appears to influence the effective scaling regime
governing lubrication-mediated motion. The present observations suggest that
the conventional Bond-number classification alone may not fully describe the
scaling behavior of lubrication-mediated motion. This viewpoint may offer a
useful framework for understanding lubrication-controlled transport of
deformable drops and bubbles in other liquid-liquid systems.

\section{Conclusions}

We have experimentally investigated the motion of water droplets sliding
beneath an inclined surface immersed in a viscous oil. The droplet velocity
follows a power-law dependence on droplet size, but the corresponding exponent
evolves systematically with inclination angle, increasing from a value close
to the quasi-spherical prediction of 3/2 toward the pancake prediction of 9/4.

Remarkably, the evolution of the velocity scaling is substantially more
pronounced than that of the macroscopic droplet shape. While the velocity data
progressively approach the pancake scaling law at large inclinations, the
global droplet dimensions exhibit a more complex evolution than expected from
a simple pancake-shape picture.

These results identify inclination angle as an important control parameter
governing lubrication-mediated droplet motion and suggest a separation between
global shape evolution and the local dissipation dynamics responsible for
droplet transport.


\textit{This work was supported by JSPS KAKENHI Grant Number JP24K00596.}


\end{document}